\documentclass[twocolumn]{aastex63}

\shorttitle{An alternative to a cosmological constant?}
\shortauthors{Loeve et al.}

\begin{document}

\title{Consistency analysis of a Dark Matter velocity dependent force
as an alternative to the Cosmological Constant}

\correspondingauthor{Steen H. Hansen}
\email{hansen@nbi.ku.dk}


\author{Karoline Loeve, Kristine Simone Nielsen \& Steen H. Hansen}
\affiliation{Dark Cosmology Centre, Niels Bohr Institute, University
  of Copenhagen, Jagtvej 128, 2100 Copenhagen, Denmark}










\begin{abstract}
A range of cosmological observations demonstrate an accelerated
expansion of the Universe, and the most likely explanation of this
phenomenon is a cosmological constant.  Given the importance of
understanding the underlying physics, it is relevant to investigate
alternative models.  This article uses numerical simulations to test the
consistency of one such alternative model. Specifically, this model
has no cosmological constant, instead the dark matter particles have an
extra force proportional to velocity squared, somewhat reminiscent of
the magnetic force in electrodynamics. The constant strength of the force is
the only free parameter.  Since bottom-up
structure formation creates cosmological structures whose internal
velocity dispersions increase in time, this model may mimic the
temporal evolution of the effect from a cosmological constant.  
It is shown that models with force linearly proportional to
internal velocites, or models proportional to velocity to power three
or more cannot mimic the accelerated expansion induced by a
cosmological constant. However, models proportional to velocity
squared are still consistent with the temporal evolution of a Universe
with a cosmological model.
\end{abstract}



\section{Introduction} \label{sec:intro}

Electromagnetism, dark matter and dark energy vary both in years since
each were discovered, and in the establishment of their theoretical 
foundations and interpretations.
Electromagnetism was discovered by Coulomb, {\O}rsted and many others
more than 200 years ago, and finally described definitively by Maxwell
150 years ago. And possibly the only remaining question is why Nature
chose at small energies to have a gauge group which involves the
$U(1)$ of the photon.

Dark matter  (DM), on the other hand, was discovered gravitationally by
Lundmark, Oort and Zwicky about 80-90 years ago, and today we still have very little
understanding of the particle properties of the dark matter, except
for upper bounds on various parameters. Dark matter 
is likely an essentially cold particle \citep{kopp2018}, 
which is very easy to envisage
from a particle physics point of view. The number of dark matter
candidates, which have been proposed, is enormous, and these models
cover a very wide range of masses, charges and gauge groups
\citep{bertone}.

The accelerated expansion of the Universe was established through the
observations of supernovae \citep{perlmutter, riess}, and subsequently
these observations have been confirmed through cosmic microwave
background and baryonic acoustic oscillations \citep{komatsu, percival,
  blake}.  A range of analyses have demonstrated that the
interpretation, that the acceleration is induced by a cosmological
constant, is in good agreement with data \citep{larson, percival,
  hicken, blake2}.

This article investigates an alternative to the cosmological constant.
Instead of the standard model with a cosmological constant and dark
matter which only has gravitational interaction, this model has no
cosmological constant. Instead the dark matter in this model has, in
addition to the gravitational attraction, a repulsive force which
depends on the dark matter velocity dispersion.  This is somewhat
similar to the Lorentz force of electromagnetism which depends on
velocity squared, in the sense that the one moving particle creates a
magnetic field, and the other particle feels a force proportional to
its own velocity times the magnetic field.  Since structure formation
proceeds bottom-up, implying that later structures have larger
internal velocity dispersions, this model may mimic the temporal
evolution of the effect of a cosmological constant.  Numerical
simulations are used to show that the observational data are still not
sufficiently detailed to reject models where the repulsive force
depends on the square of the DM velocities.

\section{The acceleration equation}
A particle at the edge of a homogeneous sphere with
radius $R(t)$ feels the gravitational acceleration, $GM/R^2$, where $M$
is the mass inside the sphere.  Here the physical radius $R(t)$ is related
to the comoving distance, $r$, through $R(t) = a(t) \, r$, where $a(t)$ is
the time-dependent scale factor.

In this description \citep{harrison1965, peebles} one can express the acceleration of that particle in the real expanding Universe
through
\begin{equation}
\frac{\ddot R}{R} = \frac{-H_0^2}{2} \left( \Omega_{m,0} \frac{r^3}{R^3} - 2 \Omega_{\Lambda,0}\right) \, .
\label{eq:rddnormal}
\end{equation}
Here and below the contribution from relativistic particles is ignored, and $\Omega_{m, 0}$ and  $\Omega_{\Lambda, 0}$  represent the
values of the densities today.

The first term on the right hand side of eq.~(\ref{eq:rddnormal}) 
corresponds to the force from the gravitational attraction of all
the matter inside a sphere of radius $R$, and the second term corresponds to a
gravitational repulsion from a cosmological constant inside a sphere of radius $R$
(the factor $-2$ arises from the $(1+ 3\omega_\Lambda)$ in the acceleration equation, since $\omega_\Lambda=-1$).
Knowing the present values of the total matter,
$\Omega_{m,0} \approx 0.3$,  cosmological constant, $\Omega_{\Lambda,0} \approx 0.7$,
and Hubble parameter, $H_0 \approx 70 km/(s\,Mpc)$,
allow us to calculate the expansion of the Universe as a function of time.

In order to see the detailed evolution we normalize with the
corresponding acceleration from a matter-only Universe, 
\begin{equation}
\frac{\ddot R_{\rm m \, only}}{R} = \frac{-H_0^2}{2} \left( \Omega_{m,0} \frac{r^3}{R^3}\right) \, ,
\end{equation}
which gives us

\begin{equation}
\frac{\ddot R}{\ddot R_{\rm m \, only}} = 1 -  2 \, \frac{\Omega_{\Lambda , 0}}{\Omega_{m,0}} \, a^3(t)  \, .
\label{eq:rddnormalnorm}
\end{equation}

This ratio evolves from unity in the early matter dominated universe, crosses zero a
few billion years ago, and is negative today where $a_{\rm today}=1$.

\section{An alternative model}
One could in principle consider an expression similar to 
eq.~(\ref{eq:rddnormalnorm}) where the time dependent part could
arise from something different from a cosmological constant
\begin{equation}
\frac{\ddot R_{\rm alternative}}{\ddot R_{\rm m \, only}} = 1 -  K(t)  .
\end{equation}
If the DM for instance has a charge which was zero at early times and
increases as the universe evolves, then a term, $K(t)$, could be
constructed such that the temporal evolution would follow that of the
normal cosmological constant term. The corresponding force would be
of the $1/r^2$ type, just like a Coulomb force.

\begin{figure}
        \includegraphics[angle=0,width=0.49\textwidth]{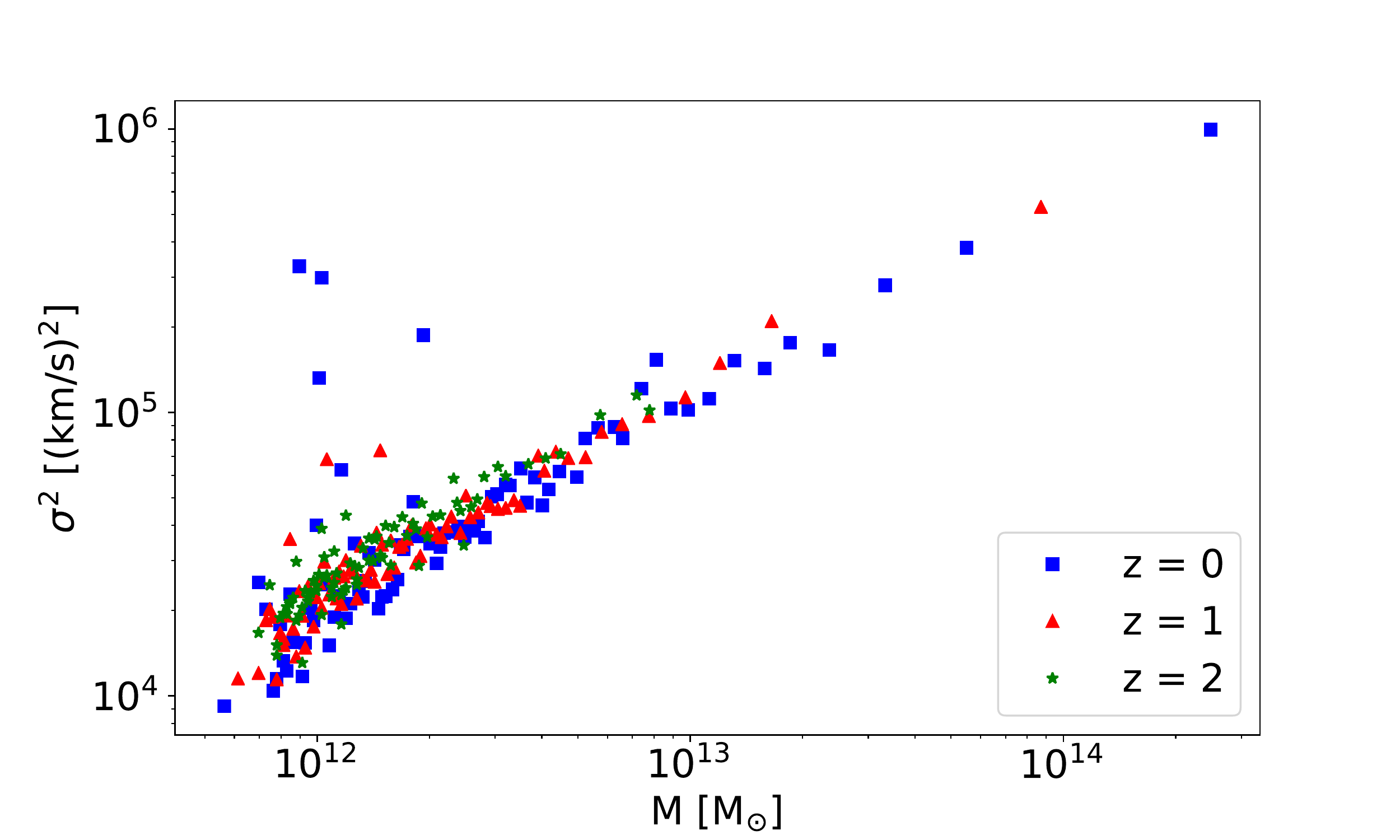}
\caption{The connection between halo mass and velocity
  dispersion. Structure formation proceeds bottom-up, which implies
  that at later times the structures have larger masses and hence higher
  velocity dispersions.}
\label{fig:fig1}
\end{figure}

This article considers a new $1/r^2$ force between DM particles. For
simplicity this force is taken to be proportional to the velocity
dispersion inside a galaxy: Imagine two galaxies at a large
distance. One DM particle in galaxy A will feel the gravitational pull
from all particles in galaxy B. And in addition, in this model,
particle A will also feel a new force proportional to $\sigma_B^2$
from all the particles in galaxy B.  The form of this force is rather
arbitrary, but it is naturally inspired by the Lorentz force which is
proportional to the (cross product) of the velocities of particles in
galaxies A and B. The only free parameter is now the constant strength
of this new force, and by choosing the sign to be negative, this force
may be repulsive.  The normalized acceleration in this universe, with
no cosmological constant but including the new force between DM
particles, now reads
\begin{equation}
\frac{\ddot R_{DM}}{\ddot R_{\rm m \, only}} = 1 + \kappa \, \sum _i \left( \frac{\sigma_i}{c}\right) ^2 \, ,
\label{eq:rddsigma2}
\end{equation}
where the individual $\sigma_i$s have been normalized to the speed of
light in order to have $\kappa$ dimensionless. It is clear, that if a
typical $\sigma$ is about $200$km/s (see figure~\ref{fig:fig1})
and one want this new term to be
of the order $-2 \times 0.7/0.3$ (see eq.~(\ref{eq:rddnormalnorm})), then
$\kappa$ must be of the order $-10^6$. 
If we compare with the forces of electromagnetism, then this model
does not have any normal charge, and hence no Coulomb force.
In figure~\ref{fig:fig1} we also see an indication of the temporal evolution, where 
typical velocity dispersions (along with the masses of the structures)
increase as the universe evolves.
The interesting question is now, what is the temporal evolution of this term
in eq.~(\ref{eq:rddsigma2}).  In order to
address this question we turn to numerical simulations.

\section{Numerically simulated universe}

The initial conditions have been produced with MUSIC \citep{MUSIC},
using cosmological parameters as observed by Planck \citep{planck}.
The numerical code RAMSES \citep{RAMSES} is used to perform a set of
pure DM simulations with different initial conditions.  With a
boxlength of 96Mpc and $2.1 \times 10^6$ particles, the individual
particle masses are $1.6 \times 10^{10} M_\odot$ .  The HOP Halofinder from
the yt Project \citep{yt, ytarticle} is used to identify structures.

For each of the 50 snapshots between redshift 20 and zero, 64
independent spheres of 9.6 Mpc comoving radii are selected. For each
sphere the acceleration felt by a particle at a random position on
that sphere is averaged, from each of the DM haloes inside the sphere,
according to eq.~(\ref{eq:rddsigma2}). This includes both the
gravitational attraction and the repulsive effect of acceleration due
to the new DM force.

\begin{figure}
        \includegraphics[angle=0,width=0.49\textwidth]{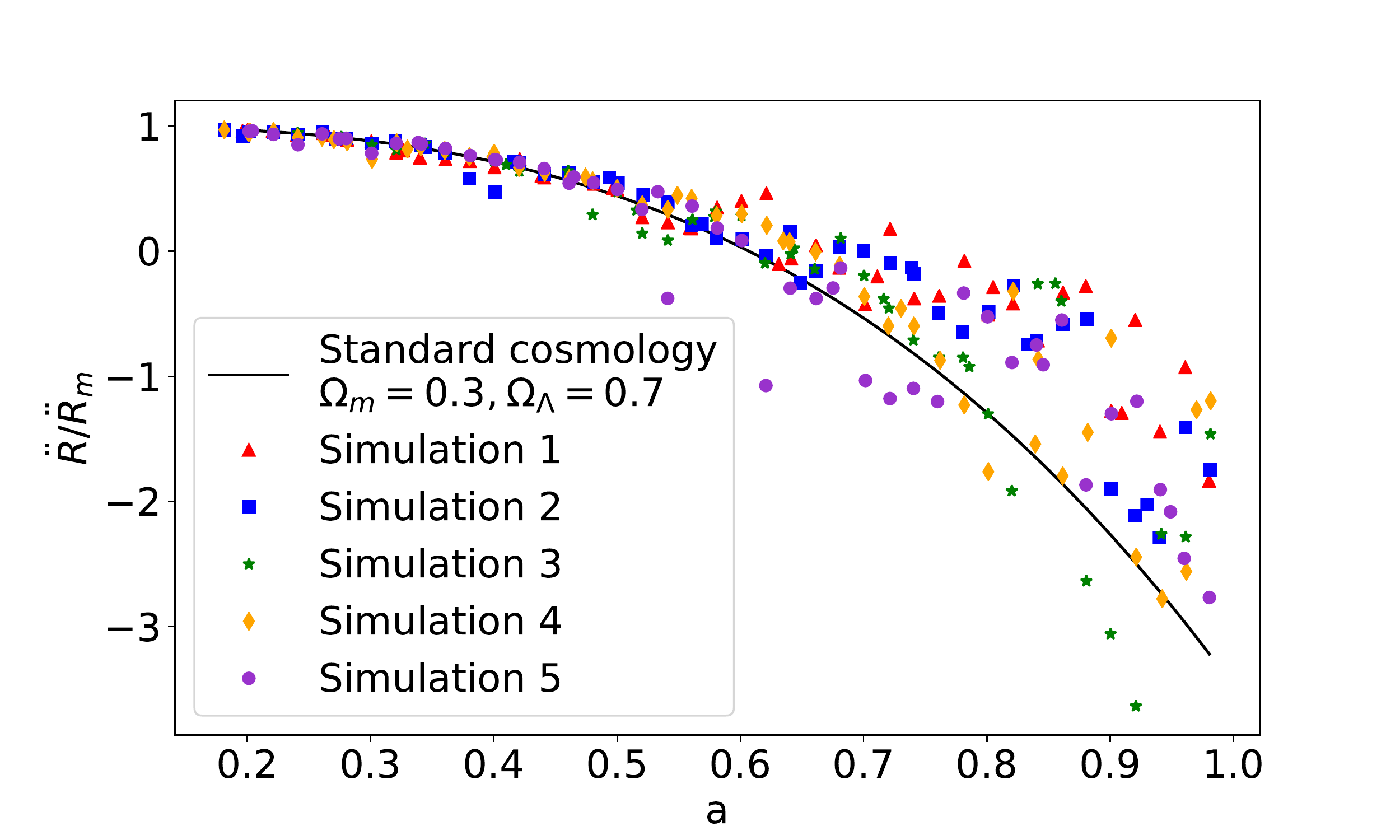}
\caption{The acceleration normalized to that of a matter dominated
  universe.  The solid line is the analytical result of a standard
  cosmology including DM and a cosmological constant.  The calculated
  symbols show the acceleration from the new force induced by
  the velocity dispersions of DM in halos.  The 5 numerical simulations
  have different initial conditions, and hence the scatter reflects the
  cosmic variance. For each simulation the magnitude of the force is
fitted in the range $0.2 <a< 0.6$. }
\label{fig:fig2}
\end{figure}

In order to determine the value of $\kappa$ the acceleration equation,
eq.~(\ref{eq:rddsigma2}) is fit to the analytical behaviour in a
standard $\Lambda$-dominated universe, eq.~(\ref{eq:rddnormal}), between
redshift 20 and $z=0.6$. The latter value is chosen since this
represents the transition between a matter dominated and a
cosmological dominated universe. If the model described above should
be able to mimic the observed acceleration of the Universe, then the
entire curve, from redshift 20 to $z=0$ should have the same shape as
the analytical shape from a Lambda-dominated universe. In figure
\ref{fig:fig2} is shown the temporal evolution of 5 numerical
simulations with different random seeds for the initial condition. As
is clear from this figure, the model has a temporal evolution which
mimics a cosmological constant fairly well.

\begin{figure}
        \includegraphics[angle=0,width=0.49\textwidth]{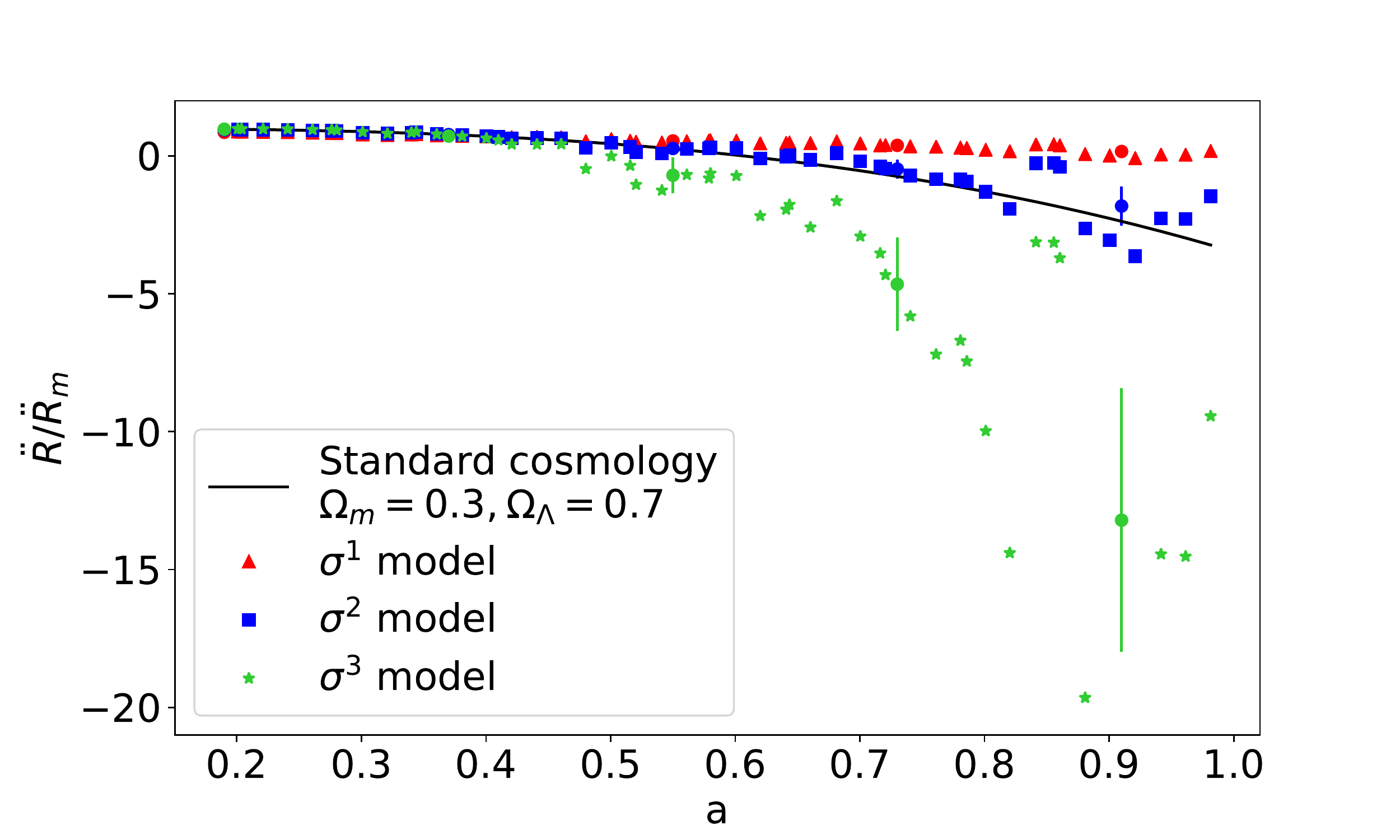}
\caption{The acceleration normalized to that of a matter dominated
  universe.  The solid line is the analytical result of a standard
  cosmology including DM and a cosmological constant.  The calculated
  symbols show the acceleration from the new force induced by
  the velocity dispersions of DM in halos.  The different symbols
reflect different possible dependencies of how the new force depends
on the velocity dispersion, 
and each set of symbols is
fitted in the range $0.2 <a< 0.6$. It is clear from the figure, that only
forces proportional to $\sigma^2$ have a temporal evolution which 
approximately mimics
that of a cosmological constant.}
\label{fig:fig3}
\end{figure}

If one instead had postulated a force which is linearly proportional
to $\left| \sigma \right|$, or to the third power, $\left| \sigma
\right| ^3$, then the temporal evolution would be significantly
different from that of a cosmological constant. This is seen in figure
\ref{fig:fig3}, where the plotted error-bars represent a spread
calculated from scatter between simulations with different initial
conditions.

\section{Limitations of the model}

The numerical simulations in this article have been performed using a
standard cosmological model.  As was shown in figure 2, the expansion
of the universe with the new DM model would rather accurately follow
this behaviour. This means that the model, within the scatter of the
numerical simulations, is consistent with the evolution driven by a
cosmologial constant model. Naturally, it would be very interesting to
modify the numerical code to correctly calculate the acceleration from
the new force, with no reference to the cosmological constant model. That would
also allow for local variations from the uniform acceleration induced by a
cosmological constant.

If the model discussed in this article should have any relevance
to Nature, then one would want to measure the {\it difference} between the new DM
force and the effect of a standard cosmological constant. That
difference will be visible in figure~2, where a departure from the
the analytical line represents a variation with respect to the cosmological
constant. We will leave it to more sophisticated simulations to 
investigate this. Similarly, one should expect that the temporal evolution
in the future should level out at a constant normalized acceleration for
this new DM force, 
because in the far future structure formation effectively ends,
leading to no increased acceleration in this model.
This is in stark contrast to the evolution from a cosmological constant,
which keeps accelerating the expansion rate.

A very important question is, to which degree other cosmological
  observables already exclude the simple model discussed above,
  including the results from BAO, CMB and SN1a observations. The
  simple answer is, that to first degree there is no difference at
  all. All these observables are derived under the assumption of the
  universe containing a distribution of photons, baryons, leptons and
  dark matter, in exactly the same way as the model described
  above. In addition the standard cosmological model used to derive
  these observables also include the changed expansion rate of the
  universe, as induced by the cosmological constant. However, this
  acceleration is typically included in the calculations by going from
  proper to comoving coordinates, and at the same time the scale
  parameter, $a(t)$, is calculated from the background cosmological
  model. To the extent that the data-points in Fig.~2 above, agree
  with the analytical derivation from the standard cosmological model
  (solid line in Fig.~2), then the scale parameter is
  unchanged. Within the scatter from the simulations with different
  initial conditions, there is perfect consistency between the
  standard cosmological model and the model discussed above, as seen
  in Fig.~2.  Specifically will all the linear properties derived in
  the standard cosmological model, including the CMB power spectra, be
  identical to the ones calculated in the model we discuss in this
  paper.  The only potentially serious problem with this model we
  discuss it at small scales.
Using a magnitude of the new DM force, which is sufficiently big to
explain the accelerated expansion of the Universe, probably leads to
inconsistencies with the internal stability of dwarf galaxies,
galaxies and galaxy clusters, since the individual DM particles are
moving with such high velocities, that the repulsive force locally may
be much larger than the gra\~vi\~tational attraction today. This problem
could be circumvented by some screening-mechanism, which should render
this new force negligible on scaller smaller than few Mpc.  This would
have negligible influence on the accumulated effect on large
scales. In addition, infalling and merging processes will continue as
in the standard picture.

The simulations presented above are limited in number of particles and 
boxsize,
and both affect the halo distribution. To test the affect of the highest
mass structures we perform a simulation with $16.8 \times 10^6$ particles
and boxlength of 192Mpc. When calculating the normalized acceleration
we find that there is virtually no difference from the smaller 
boxsize simulation because of the very sharp
drop in the halo-mass-function at high masses.
Similarly we  perform a simulation with $16.8 \times 10^6$ particles
and boxlengths of 96 Mpc, in order to see the effect of the many
smaller structures. Again we see that there is a very small effect, because
the $M-\sigma^2$ variation is more important than the increase in 
number of smaller haloes. Finally we have assumed that the
new force is proportional to the velocity squared, summed over all
haloes. To this end we are implicitly assuming that first identifying
the haloes (including all particles which have departed from the general 
expansion of the universe), and then subsequently summing over all haloes,
is the correct approach. We wish to consider alternatives to this
approach in a future study.

\section{Conclusion}
With a purely phenomenological approach, we allow the dark matter
particles to have a new repulsive force proportional to the squared
internal velocity dispersions of cosmological halos. Since
structure-formation proceeds bottom-up, this implies that this force
grows as a function of time. We use numerical cosmological simulations
to show that this force may mimic the temporal evolution of a
cosmological constant. We also find that similar forces linearly
proportional to the velocity dispersion, or to power three, are
inconsistent with the temporal evolution of our Universe.

\acknowledgments
We thank Davide Martizzi for support with all numerical aspects, including
running RAMSES.
This project is partially funded by
the Danish council for independent research, DFF 6108-00470.







\end{document}